\documentclass{aa501}

\usepackage{graphicx}

\usepackage{graphicx,epsfig}
\def\arcsec{$^{\prime\prime}$}
\def\arcmin{$^{\prime}$}
\def\degrees{$^{\circ}$}
\def\etal{\rm et al.~}

\begin{document}

\title{Radio and X-ray diffuse emission in six clusters
of galaxies}
\author{F. Govoni\inst{1,2} \and L. Feretti\inst{2}
G. Giovannini\inst{2,3} \and  H. B\"ohringer\inst{4} \and
T. H. Reiprich\inst{4} \and M. Murgia\inst{2,3}}

\offprints{F. Govoni (fgovoni@ira.bo.cnr.it)}

\institute{
Dipartimento di Astronomia, Univ. Bologna, Via Ranzani 1, I--40127 Bologna, Italy
\and Istituto di Radioastronomia -- CNR, via Gobetti 101, I--40129
Bologna, Italy
\and Dipartimento di Fisica, Univ. Bologna,  
Via B. Pichat 6/2, I--40127 Bologna, Italy
\and Max Planck Institut f\"ur Extraterrestrische Physik, PO Box 1603, 
D--85740 Garching, FRG}

\date{Received ; accepted}

\abstract{
Deep Very Large Array radio observations confirm the presence of halo and
relic sources in six clusters of galaxies (A115, A520,
A773, A1664, A2254, A2744) where a wide diffuse emission
was previously found in the NRAO VLA Sky Survey.
New images at 1.4 GHz of these six clusters of galaxies are presented  
and X-ray data obtained from the ROSAT archive are analyzed.
The properties of clusters hosting radio halos and relics are analyzed and
discussed. A correlation between the halo radio power and the cluster 
gravitational mass is presented.
\keywords{
Galaxies: clusters : general -- intergalactic medium -- magnetic fields
-- Radio continuum: general -- 
X-rays: general
}
}

\maketitle

\section{Introduction}

An important problem in cluster phenomenology regards cluster-wide
radio halos, whose prototype is Coma C (Willson 1970).

Radio halos in clusters of galaxies are diffuse radio sources
located in the cluster center with no obvious connection to
the galaxy population of the clusters.
They are characterized by a typical size of about 1 Mpc, regular shape,
low surface brightness and steep radio spectrum.
Other similar sources, but in general with irregular morphologies, are
named relics. They are found at the cluster periphery.
In a few clusters, both a central halo and a peripheral relic are
present.

These radio sources demonstrate
the existence of relativistic electrons and large scale magnetic fields
in the cluster intergalactic medium (IGM).

The study of these sources is very important since they are large
scale features, which are related to other cluster properties in the
optical and X-ray domain, and are thus directly connected to the
cluster history and evolution.

Radio halos and relics have always been considered very rare
structures. However, thanks to the sensitivity of the radio
telescopes and to the existence of deep radio surveys,
the number of known halos and relics has recently increased.

Giovannini \etal (1999),
using the NRAO Very Large Array (VLA) 
Sky Survey (NVSS, Condon \etal, 1998),
searched for new radio halos and relics in the sample
of X-ray-brightest Abell cluster (XBACs)
presented by Ebeling \etal (1996).

The XBACs sample consists of clusters/subclusters from 
the catalogue of Abell et al. (1989) detected in the 
ROSAT All Sky survey (RASS) with X-ray flux f$_{\rm X}>$ 5 10$^{-12}$
erg cm$^{-2}$ s$^{-1}$ in the 0.1-2.4 keV energy range.
This is an all-sky, X-ray flux limited sample, 
complete in the galactic latitude range $\mid b \mid
\ge$20\degrees~ and in the redshift interval z$\le$0.2, but 
it contains also  12 clusters at lower galactic latitude and 24 clusters 
with redshift greater than 0.2 that meet the flux criterion.
The NVSS  was performed at 1.4 GHz with the VLA 
in D configuration. It has an angular resolution of 45\arcsec~ (FWHM),
a noise level of 0.45 mJy/beam (1$\sigma$) and covers all the sky north of 
Declination = --40\degrees.
Because of the lack of short baselines, Giovannini \etal (1999)
used as a limiting redshift for the search of diffuse 
sources the value z$\geq$0.044. 
Considering this redshift limit, and
taking into account the declination limit of the NVSS, they
searched for diffuse emission in 207 clusters.
They found 29 clusters, 11 of which were already known in 
the literature to possess a radio halo or relic, whereas 18 were new 
detections.

To properly map these diffuse sources high sensitivity
to the extended features is needed, but also a good resolution is
necessary to distinguish a real diffuse source from the blend of
unrelated sources.

We started a project
to study all these new radio halo and relic candidates
with improved sensitivity and resolution
with respect to the NVSS.
Here we present the data for the clusters A115, A520, A773, A1664, A2254, 
A2744 which were included in the first run of observations.
The remaining clusters are presently being
analyzed and will be presented in a future work.

The similarity between the radio and the X-ray morphology, 
the high cluster X-ray temperature and luminosity, 
and the evidence of substructures in clusters with 
radio halos (e.g. Feretti 1999) 
indicate the importance to analyze the X-ray properties 
of the cluster containing these diffuse sources.
To obtain X-ray information
and to find correlations between the radio and the X-ray parameters
in clusters with radio halos and/or relics we analyzed 
pointed ROSAT observations for these clusters.

Here we present new radio images of these clusters, and
compare them with the X-ray emission. 

In Sect. 2 we present the radio and the X-ray data. In Sect. 3 we give
the radio and the X-ray images. In Sect. 4 we perform
an X-ray analysis, while the discussion and 
conclusions are reported in Sect. 5.

We use a Hubble constant H$_0$=50~km~s$^{-1}$Mpc$^{-1}$ and q$_0$=0.5
throughout the paper.

\begin{table}
\caption{General properties of the clusters}
\begin{flushleft}
\begin{tabular}{ccccc}
\hline
\noalign{\smallskip}
Name      &  z            & Richness & BM type  & kpc/$''$  \\
          &               &          &          &           \\
\noalign{\smallskip}			                   
\hline		    			                   
\noalign{\smallskip}			                   
 A115     & 0.1971        & 3        & III      & 4.18      \\
 A520     & 0.199         & 3        & III      & 4.21      \\
 A773     & 0.217         & 2        & II-III   & 4.47      \\
 A1664    & 0.1276        & 2        &          & 3.00      \\
 A2254    & 0.178         & 2        &          & 3.88      \\
 A2744    & 0.308         & 3        & III      & 5.58      \\
\noalign{\smallskip}
\hline
\multicolumn{5}{l}{\scriptsize Col. 1: cluster name; Col. 2: redshift;
Col. 3: richness class;}\\
\multicolumn{5}{l}{\scriptsize Col. 4:  Bautz \& Morgan classification;}\\ 
\multicolumn{5}{l}{\scriptsize Col. 5: linear to angular conversion.}\\
\end{tabular}
\end{flushleft}
\end{table}

\section{Data}

Some general properties of the six clusters of galaxies analyzed here 
are given in Table 1. 
Deep radio images were obtained for all these clusters.
The X-ray emission was analyzed using
the data taken from the ROSAT public archive.
The presence of sub-structures in X-ray images
can give indications on the dynamical evolution
of the clusters, however we also searched in literature 
for optical data indicating 
peculiar sub-structures in the galaxy distribution 
that could be interpreted in terms of cluster merging.
As reported in the individual notes of the clusters, we found
useful information for A115 and A520.

\subsection{Radio observations}

\begin{table}
\caption{VLA Observing Log}
\begin{flushleft}
\begin{tabular}{ccccc}
\hline
\noalign{\smallskip}
Name &  Frequency  & Bandw.& Config. & Date  \\
     & MHz  &   MHz        &         &       \\
\noalign{\smallskip}
\hline
\noalign{\smallskip}
 A115     & 1365/1435     & 50   & C & December 98 \\
          & 1365/1665     & 50   & D & March 99    \\
 A520     & 1365/1435     & 50   & C & December 98 \\
          & 1365/1665     & 50   & D & March 99    \\
 A773     & 1365/1435     & 50   & C & November 98 \\
          & 1365/1665     & 50   & D & March 99    \\
 A1664    & 1365/1435     & 50   & C & November 98 \\
 A2254    & 1365/1435     & 50   & C & December 98 \\
          & 1365/1665     & 50   & D & April 99    \\
 A2744    & 1365/1435     & 50   & C & November 98 \\
          & 1365/1665     & 50   & D & March 99    \\
\noalign{\smallskip}
\hline
\multicolumn{5}{l}{\scriptsize Col. 1: cluster name; Col. 2: observing frequency;}\\
\multicolumn{5}{l}{\scriptsize Col. 3: bandwidth; Col. 4: VLA configuration;}\\ 
\multicolumn{5}{l}{\scriptsize Col. 5: observing date. }\\
\end{tabular}
\end{flushleft}
\end{table}

\begin{table}
\caption{Primary flux density calibrators and phase calibrators}
\begin{flushleft}
\begin{tabular}{ccc}
\hline
\noalign{\smallskip}
Name &  Primary Cal.  & Phase Cal. \\
\noalign{\smallskip}
\hline
\noalign{\smallskip}
 A115     & 3C48     & $0119+321$  \\
 A520     & 3C48     & $0459+024$  \\
 A773     & 3C286    & $0903+468$  \\
 A1664    & 3C286    & $1248-199$  \\
 A2254    & 3C286    & $1745+173$  \\
 A2744    & 3C48     & $0025-260$  \\
\noalign{\smallskip}
\hline
\label{tabcal}
\end{tabular}
\end{flushleft}
\end{table}

The radio data  were obtained with the VLA
at 1.4 GHz, in the C and D configurations for a total
observing time of about 2 hours for each cluster in 
each configuration.
The observing frequency and configurations 
were chosen in order to have a good
sampling of short spacings, 
ensuring that a halo-type source could be easily detected and imaged.

The shortest baseline is 35 m, corresponding to $\simeq$175$\lambda$.
Therefore structures up to about 20\arcmin~ in angular size
are fully imaged.
We note that our sources are all smaller in size.
Only for A1664 where D array observations are not available
we have a lack of short baselines which can affect our image of the
relic radio source as discussed in Sect. 3.4.

\begin{table*}[t]
\caption{X-Ray data from the ROSAT archive}
\begin{flushleft}
\begin{tabular}{cccccc}
\hline
\noalign{\smallskip}
Name      &  Detector & texp.   & RA        & DEC       & ROR\\
          &             & (sec) & (J2000.0) & (J2000.0) &
\\
\noalign{\smallskip}
\hline
\noalign{\smallskip}
 A115     & HRI         & 29692 & 00 55 50.04 & 26 24 36.0  & 800633h-1
\\
 A520     & PSPC        &  4850 & 04 54 43.02 & 02 52 12.0  & 800480p
\\
 A773     & HRI         & 16663 & 09 17 52.08 & 51 43 48.0  & 800618h
\\
 A1664    & PSPC        & 12760 & 13 03 43.02 & $-$24 15 00.0 & 800411p
\\
 A1664    & HRI         & 22227 & 13 03 43.02 & $-$24 15 00.0 & 800749h
\\
 A2254    & HRI         & 21976 & 17 17 50.0  &  19 40 12.0 & 800882h-1
\\ 
 A2744    & PSPC        & 13648 & 00 14 19.02 & $-$30 23 24.0 & 800343p
\\
\noalign{\smallskip}
\hline
\multicolumn{6}{l}{\scriptsize Col. 1: cluster name; Col. 2: ROSAT detector; Col. 3: exposure time;}\\
\multicolumn{6}{l}{\scriptsize Col. 4, Col. 5: pointing position; Col. 6: Rosat Observation Request Sequence Number.}\\ 
\end{tabular}
\end{flushleft}
\end{table*}

\begin{table*}
\caption{Radio parameters}
\begin{flushleft}
\begin{tabular}{ccccccccc}
\hline
\noalign{\smallskip}
Name  &type    & $S_{1400}$    & LLS          & $\alpha$ &
P$_{1400}$     & P$_{\rm tot}$   & Volume       & H$_{\rm eq}$ \\
      &        & mJy         &    Mpc       &          &
Watt/Hz        & Watt        & Mpc$^3$      & $\mu$G  \\
\noalign{\smallskip}
\hline
\noalign{\smallskip}
 A115 & R  &14.7     &  2.5   &  1.1   & $2.7\times 10^{24}$ &$3.1\times 10^{34}$ & 0.33 &  0.6  \\
 A520 & H  &34.4     &  1.4   &  1.7   & $6.4\times 10^{24}$ &$4.0\times 10^{35}$ & 0.42 &  1.3  \\
 A773 & H  &12.65    &  1.6   &  2.8   & $2.8\times 10^{24}$ &$1.6\times 10^{37}$ & 0.32 &  4.3  \\
 A1664& R  &50.2     &  1.1   &  (1.2) & $3.7\times 10^{24}$ &$5.2\times 10^{34}$ & 0.17 &  0.8  \\
 A2254& H  &33.7     &  1.2   &  1.2   & $4.9\times 10^{24}$ &$7.0\times 10^{34}$ & 0.40 &  0.7  \\
 A2744& H  &57.1     &  2.3   &  1.4   & $2.6\times 10^{25}$ &$6.3\times 10^{35}$ & 3.40 &  0.8  \\
 A2744& R  &18.2     &  2.0   &  $>$ 2 & $8.5\times 10^{24}$ &$> 1.6\times 10^{36}$& 0.46 & $>$1.9 \\
\noalign{\smallskip}
\hline
\multicolumn{9}{l}{\scriptsize Col. 1: cluster name; Col. 2: source type (R=relic, H=halo); 
Col. 3: flux density;}\\
\multicolumn{9}{l}{\scriptsize Col. 4: largest linear size; Col. 5: average spectral index,}\\ 
\multicolumn{9}{l}{\scriptsize  Col. 6: radio power at 1400 MHz;
Col. 7: total radio power between 10 MHz to 10 GHz;}\\
\multicolumn{9}{l}{\scriptsize Col. 8: radio source volume; Col. 9: equipartition magnetic field.}\\ 
\end{tabular}
\end{flushleft}
\end{table*}

The details of the radio observations are given in Table 2.
The observations in D configuration were performed at distant frequencies
within the same band,
to obtain information on the spectral index of the diffuse sources.

The data were calibrated and reduced with the Astronomical
Image Processing System
(AIPS), following the standard procedure: Fourier-Transform, Clean and
Restore. Self-calibration was applied to minimize the effects of
amplitude and phase variations.

The source 0137+331 (3C48) or 1331+305 (3C286) was used as a primary
flux density calibrator (see Table 3).
The phase calibrator was
observed every 20 minutes, approximately.

We first reduced separately the data from the two  different
configurations, for a consistency check.
Final images were produced by adding together the C configuration
data at 1365 and 1435 MHz, with the D configuration data  at 1365 MHz.
Images were produced with different resolutions,
using the AIPS task IMAGR. The image with the
highest available resolution ($\simeq 15''$) provides  
information on the
discrete radio sources, while the low resolution map ($\simeq 50''$) 
allows the proper detection
of the low brightness diffuse radio emission.
All the data at 1665 MHz are affected by interferences.
A determination of the spectral index images 
is therefore difficult and only a total spectrum can 
be estimated.

\subsection{X-Ray data}

\begin{figure*}
\resizebox{18 cm}{!}{\includegraphics {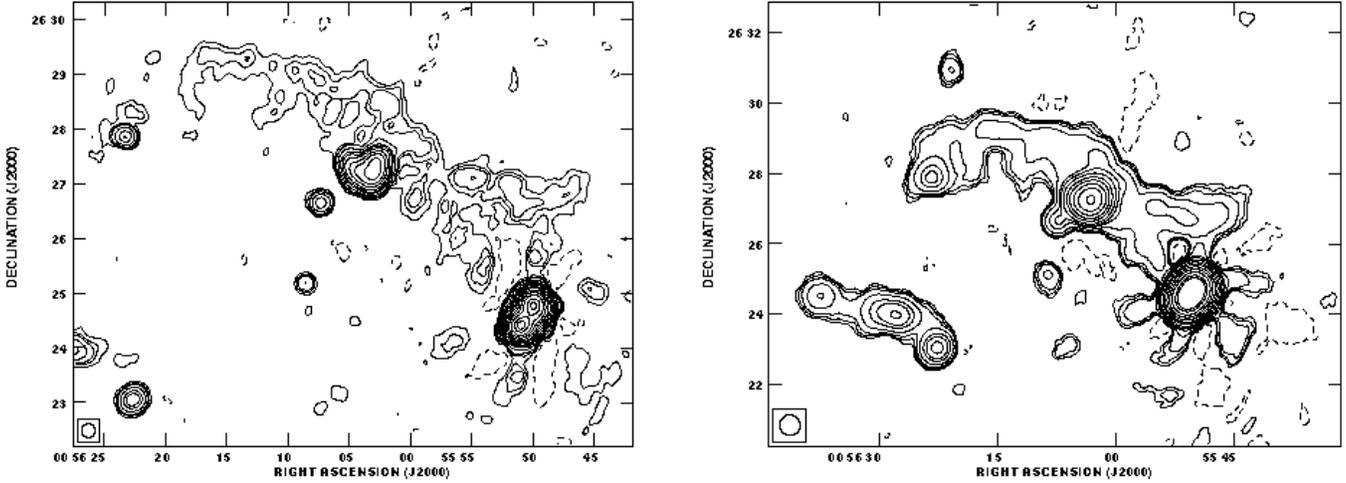}}
\caption[]{Radio images at 1.4 GHz of the relic source in A115.
Left panel: the FWHM is 15\arcsec $\times$ 15 \arcsec; 
the noise level is 0.1 mJy/beam.
Contour levels are: -0.4 0.4 0.7 1 2 4 6 10 19 32 57 97 169 292 400 mJy/beam.
Right panel: the FWHM is 35\arcsec $\times$ 35\arcsec; the noise level is 0.1 mJy/beam.
Contour levels are:
-0.5 0.5 0.7 1 2 4 6 10 19 32 57 97 169 292 400 mJy/beam.
The pattern around the strong source 3C28 
it is due to dynamic range problems. 
The peak of the image is coincident with this 
radio galaxy and it is 0.96 Jy/beam.}
\label{A115_a}
\end{figure*}

\begin{figure}
\resizebox{8 cm}{!}{\includegraphics {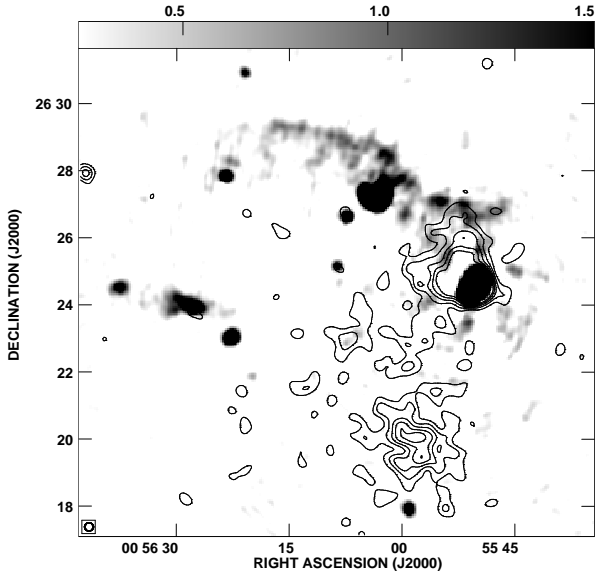}}
\hfill
\caption[]{Isophotal contour plots
of the X-ray (HRI) image taken from the ROSAT archive,
overlapped to the 1.4 GHz grey-scale image (FWHM=15 \arcsec)
of the cluster A115.
The X-ray contour levels are:
1.0 1.2 1.4 1.6 1.8 3.7 5.5 8.3 Counts/pixel
(1pixel=5\arcsec$\times$5\arcsec).
}
\label{A115_b}
\end{figure}

We used archival observations of the 6 clusters
of galaxies (A115, A520, A773, A1664, A2254, A2744) carried out
with the Position Sensitive Proportional Counter 
(PSPC) and/or the High Resolution Imager (HRI) on ROSAT satellite.

The X-ray observations are identified by their Rosat Observation 
Request Sequence Number (ROR) in Table 4, where the detector,
the exposure time, and the pointing position are also listed.
When more than one observation existed for the same
object with the same instrument, we used the observation
with the longest exposure time. 

The ROSAT energy band is 0.1-2.4 keV for both detectors. 
The HRI provides a FWHM spatial resolution on axis $\simeq$ 5$''$
and basically no spectral resolution, while the PSPC has a
moderate angular resolution (FWHM$\simeq$ 25$''$ on axis), 
as well as spectral resolution.
A detailed description of the instruments can be
found in Tr\"umper (1983) and Pfeffermann \etal (1987).

The data analysis has been performed with the EXSAS package (Zimmermann
\etal 1994).

\section{Radio and X-ray images}

In the radio images presented here, the presence
of diffuse sources is confirmed in all the six clusters of galaxies.
In Table 5 we give the relevant parameters of the radio emission
derived as follows:

1) the flux density was obtained after subtraction of the
discrete sources, by integrating the surface brightness down to
the noise level; 

2) the radio spectral index was derived using the D configuration
images at 1365 and 1665 MHz. Taking into account the 
closeness of the two frequencies, and 
some interferences in the images at 1665 MHz, this estimate
of the spectral index should be taken cautiously with a typical
error of about 0.5.
For A1664, which lacks observation at 1665 MHz, 
we assume  $\alpha=1.2$, as found in typical relic radio sources
(e.g. 1253+275 in Coma);

3) the total power between 10 MHz to 10 GHz was derived
using the spectral index given in the table;

4) the  equipartition magnetic field was evaluated assuming
a low frequency cutoff of 10 MHz, 
a high frequency cutoff of 10 GHz, equal energy density in 
protons and electrons, and a volume filling factor of 1. 
The source volume was considered to be an ellipsoid.

The images of the X-ray brightness distribution presented
in this work were produced by binning
the photon events in a two-dimensional grid and then smoothing
with a Gaussian filter. The PSPC images were produced in the hard band,
corresponding to the energy range 0.5-2.0 keV.
With the selection of this range of energies we minimize the
Galactic and particle background (which dominate below 0.5 keV 
and above 2.0 keV, respectively) while retaining most of the 
cluster emission.
In the following we present the results and comment
on each individual cluster.

\begin{figure*}
\resizebox{18 cm}{!}{\includegraphics {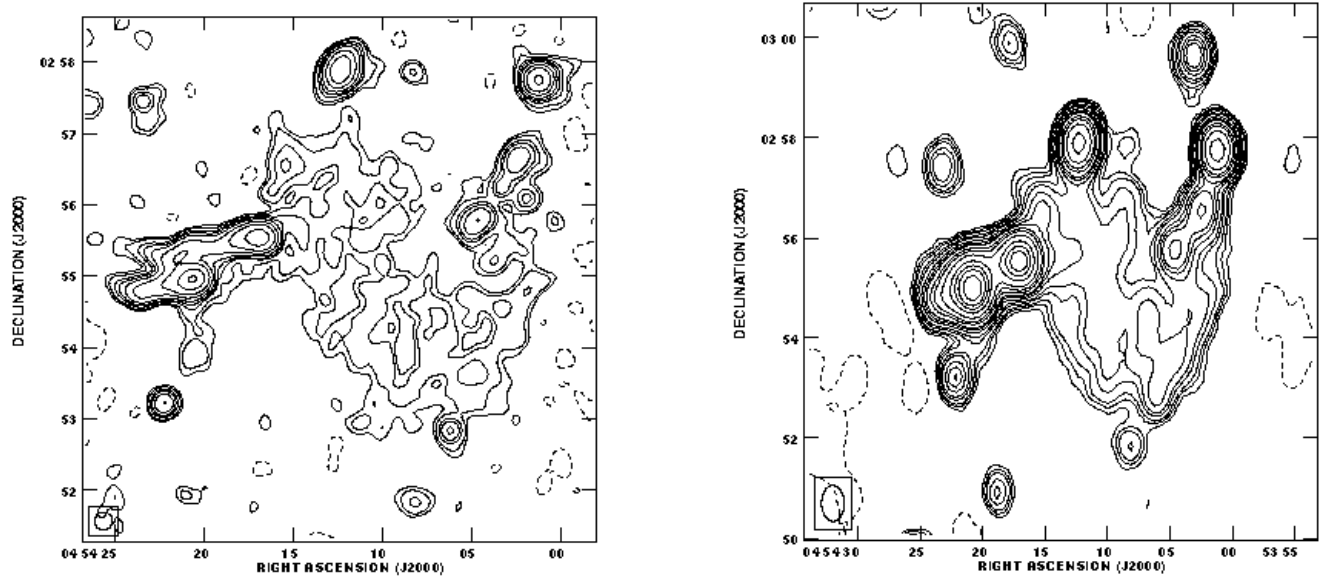}}
\caption[]{
Radio images at 1.4 GHz of the halo source in A520.
Left panel: the FWHM is 15$\times$15\arcsec;
the noise level is 0.03 mJy/beam. Contour levels are:
-0.06 0.06 0.1 0.2 0.3 0.5 1 2 5 10 20 mJy/beam.
Right panel: the FWHM is 43\arcsec $\times$27\arcsec;
the noise level is 0.03 mJy/beam. Contour levels are:
-0.1 0.1 0.15 0.2 0.3 0.4 0.6 0.8 1 1.5 2 3 5 7 9 12 18 25 mJy/beam.
}
\label{A520_a}
\end{figure*}

\begin{figure}
\resizebox{8 cm}{!}{\includegraphics {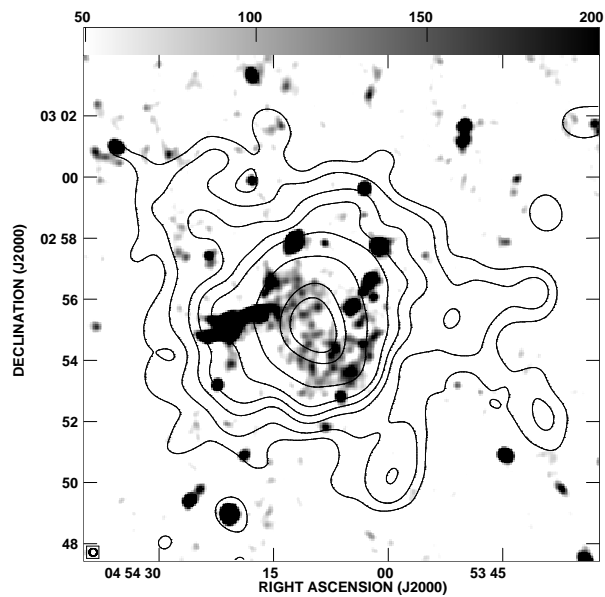}}
\hfill
\caption[]{Isophotal contour plots of the
X-ray (PSPC) image taken from the ROSAT archive,
overlapped to the 1.4 GHz grey-scale image
(FWHM 15\arcsec) of the cluster A520.
The X-ray contours are: 0.2 0.3 0.5 0.7 1 2 4 5 Counts/pixel (1pixel=15\arcsec$\times$15\arcsec). }
\label{A520_b}
\end{figure}

\subsection{A115}

The 1.4 GHz radio images obtained  with different
angular resolutions  are shown in Fig. \ref{A115_a}, where the 
diffuse emission is easily visible. 

The bright radio galaxy in the SW corner located
at about RA(2000)=00$^h$55$^m$50$^s$, DEC(2000)=26\degrees 
24\arcmin 34\arcsec~~is  3C28. This source, showing here a
double radio structure, has been studied in detail by 
Feretti \etal (1984) and Giovannini \etal (1987). In their images,
this radio galaxy shows an unusual distorted morphology. 
It is characterized by two 
components located on both sides of the optical galaxy, with low brightness 
tails in the western direction, and without a clear radio
compact core. The other strong source embedded within the diffuse emission
is the  radio source 0056+265, a head-tail 
radio galaxy located at RA(2000)=00$^h$56$^m$03$^s$,  
DEC(2000)=26\degrees 27\arcmin 13\arcsec 
(Gregorini \& Bondi 1989). Both radio sources
are associated with cluster galaxies.

The relic is elongated in the SW-NE
direction with a projected maximum extension of about 10$'$ (2.5 Mpc)
and a transverse size about 5 times smaller.
Other relic sources so extended are already
known in the literature (see e.g. A3667, R\"ottgering \etal 1997). 

In the higher resolution map, the Northern source boundary shows
a sharper edge than on the opposite side. A similar feature
is seen in other relics (e.g. 1253+275 in the Coma cluster, 
Giovannini \etal 1991; A3667, 
R\"ottgering \etal 1997).
This could be related to some compression of the plasma.

A115 is known in literature to be characterized by a double
X-ray peak (Forman \etal 1981).
The X-ray data  at our disposal have been obtained with 
the ROSAT HRI.
In Fig. \ref{A115_b} we show the X-Ray image (contours),
smoothed with a Gaussian of $\sigma$=10$''$
over-imposed onto the radio image (grey scale)
at resolution of 15$''$.
In the figure, a double X-ray morphology is evident 
with two irregular enhancements in the X-ray surface
brightness distribution. 
Part of the X-Ray emission in the northern peak could come from
the contribution of the radio galaxy 3C28.

\begin{figure*}
\resizebox{18 cm}{!}{\includegraphics {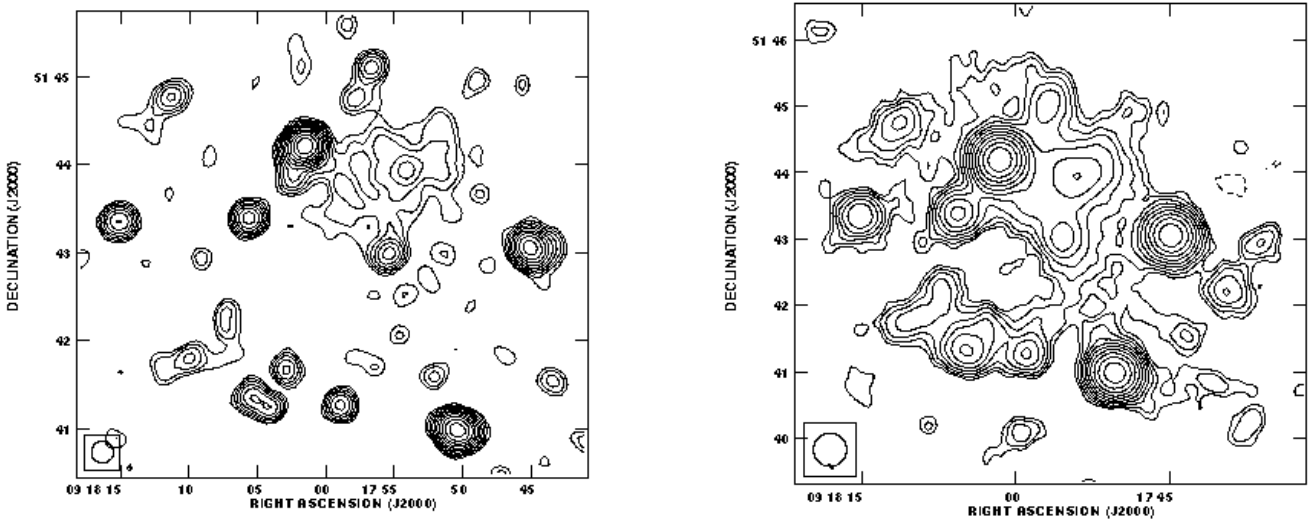}}
\caption[]{
Radio images at 1.4 GHz of the halo source in A773.
Left panel: the FWHM is 15\arcsec$\times$15\arcsec;
the noise level is 0.02 mJy/beam. Contour levels are:
-0.06 0.06 0.08 0.12 0.17 0.24 0.34 0.48 0.68 0.96 1.36 1.92
2.71 3.84 5.43 7.68 10.86 mJy/beam.
Right panel: the FWHM is 30\arcsec$\times$30\arcsec;
the noise level is 0.03 mJy/beam. Contour levels are:
-0.07 0.07 0.1 0.14 0.2 0.28 0.40 0.56 0.79 1.12 1.58
2.24 3.17 4.48 6.34 8.96 12.67 17.92 mJy/beam.}
\label{A773_a}
\end{figure*}

\begin{figure}
\resizebox{8 cm}{!}{\includegraphics {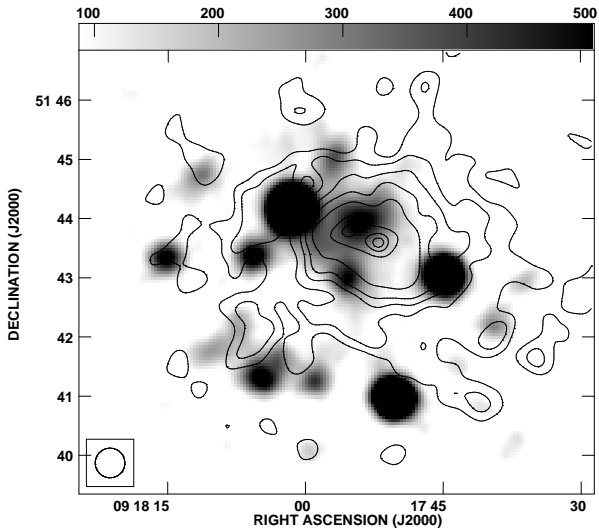}}
\hfill
\caption[]{Isophotal contour plots
of the  X-ray (HRI) image taken from
the ROSAT archive overlapped to the grey-scale
image at 1.4 GHz (FWHM  30\arcsec)
 of the cluster A773.
The X-ray contours are: 0.65 0.8 1 1.5 3 3.4 3.6
Counts/pixel (1pixel=5\arcsec$\times$5\arcsec). }
\label{A773_b}
\end{figure}

The diffuse radio source belongs to the northern X-ray clump, and
extends from the sub-cluster center to the periphery. 
According to its non-central location and its elongated structure, it
is classified as a cluster relic. 
However, elongated relics, as for example those in Coma, A3667, A2255, 
are generally at the cluster periphery, and with the largest dimension
roughly perpendicular to the cluster radial orientation.
So this source is quite unusual,
and difficult to interpret in the framework of currently
proposed models, invoking spherically symmetric
compression of the IGM by shock waves during the cluster formation
(En{\ss}lin \etal 1998, Roettiger \etal 1999).
A possibility is that this source is actually in foreground or
background with respect to the cluster center.

We have considered the possibility that this radio source
could be part of a supernova remnant, but we have found  no identification 
in the catalog of Green (2000).
Inspection of a large field of the NVSS reveals no other
diffuse radio structure, which could be part of the same 
supernova remnant.
Moreover the radio spectrum  
found for this source is much steeper than typical
supernova remnant spectra (Green 2000).
Therefore this interpretation seems very unlikely,
confirming that this source should be a cluster relic.

In the optical band, Beers \etal (1983) mapped the galaxy 
distribution of the cluster over a field of 3.2 Mpc square.
 The galaxy distribution reveals the presence of three primary 
clumps of galaxies.

They further compared the X-ray image from Forman \etal (1981) with the
optical galaxy distribution finding that the two main
optical clumps correspond to the two 
peaks in the X-ray distribution. The third optical 
clump, located at 7\arcmin~to the east of the two main peaks,
 has no X-ray emission detected. 
We note, however, that the third optical peak is located in 
coincidence of the radio elongated emission visible 
in the bottom left corner of the low resolution radio 
image at about RA(2000)=00$^h$ 56$^m$ 30$^s$, 
DEC(2000)=$+$26\degrees 24\arcmin, 
(Fig.\ref{A115_a}, right panel).

\begin{figure*}
\resizebox{18 cm}{!}{\includegraphics {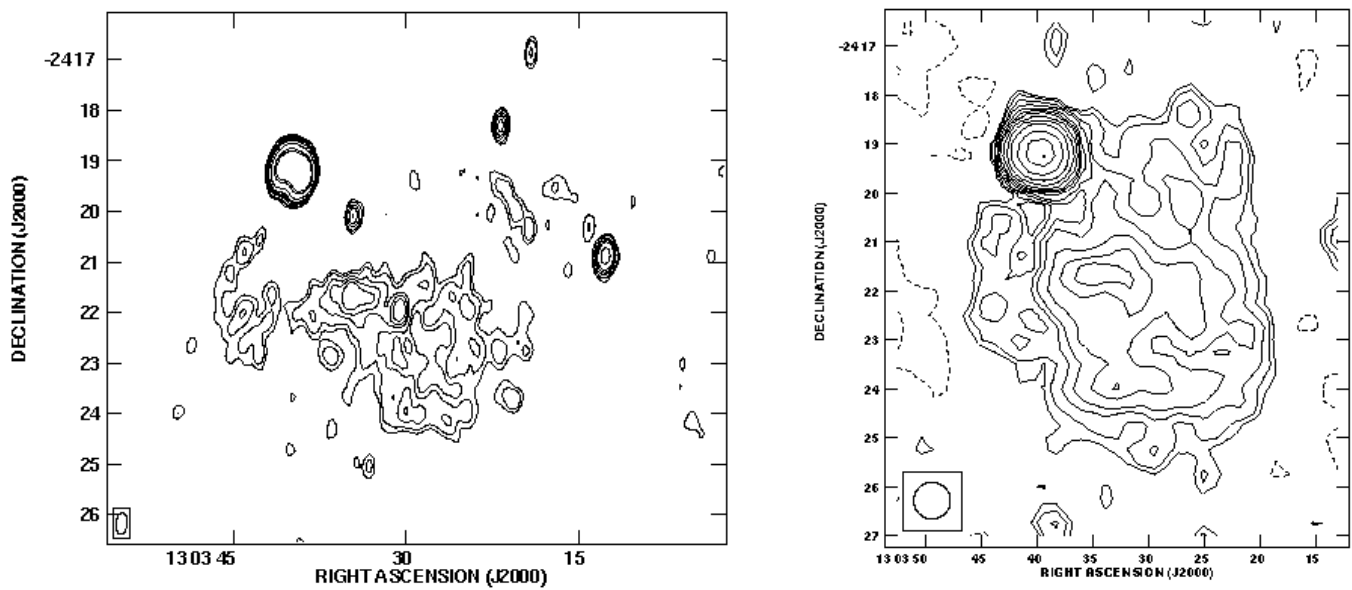}}
\caption[]{
Left panel: radio images of the relic source in A1664.
 The FWHM is 26\arcsec $\times$ 13\arcsec;
the noise level is 0.07 mJy/beam. Contour levels are:
0.2 0.3 0.5 0.7 1 2 5 7 mJy/beam.
Right panel: radio image of A1664 taken from the NVSS.
Contour levels are:
-1 0.7 1 1.5 2 3 4 5 7 10 15 25 40 55 mJy/beam.
}
\label{A1664_a}
\end{figure*}

\begin{figure}[h]
\resizebox{8 cm}{!}{\includegraphics {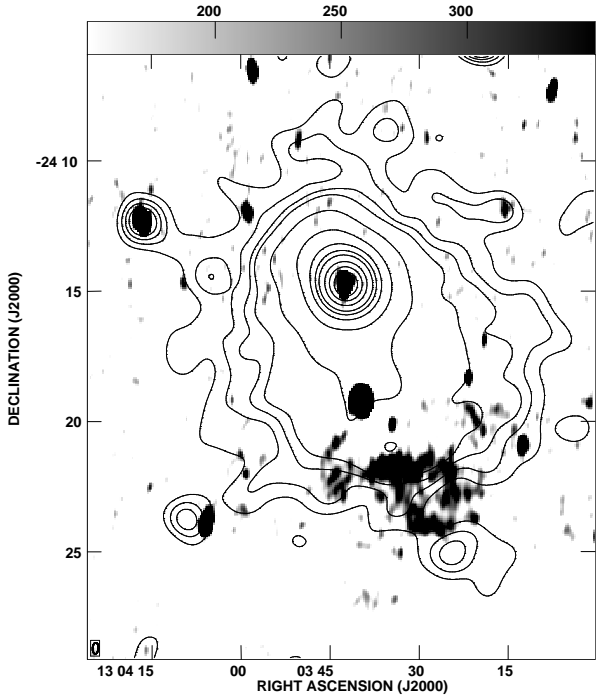}}
\hfill
\caption[]{Isophotal contour plots of the
 X-ray (PSPC) image taken from the ROSAT archive
overlapped to the grey-scale image at 1.4 GHz
(FWHM 26\arcsec $\times$ 13\arcsec) of the cluster A1664.
The X-ray contours are: 0.4 0.6 0.8 1 2 5 10 15 20 30 40
50 60 Counts/pixel (1pixel=15\arcsec$\times$15\arcsec).}
\label{A1664_b}
\end{figure}

\subsection{A520}

The 1.4 GHz images of the center of A520 are shown in Fig. \ref{A520_a}.
The radio halo shows a low surface brightness with a clumpy
structure. The structure is slightly elongated
in the NE-SW direction, with a maximum projected size of 
about 340$''$ (1.4 Mpc).

There are several discrete radio sources at the boundary 
of the halo. The two head-tail radio sources
located on the eastern side of the halo
are identified with cluster galaxies.
The projected extension is about 600 kpc  for 
the northern head-tail radio galaxy and about
400 kpc for the southern one.
It is remarkable the fact
that the tails in these sources are oriented toward the same
direction, opposite to the cluster center.
According to Bliton \etal (1998) this
could be related with  bulk motion of the 
intergalactic medium and could therefore indicate the presence of a merger
in this cluster. Alternatively they could both belong to the 
same infalling group. 

The X-ray image of this cluster, obtained with  the ROSAT PSPC,
 was produced by binning
the photon events in a two-dimensional grid and then smoothing
with a Gaussian filter of $\sigma$=30$''$.
In Fig. \ref{A520_b} we present the X-ray image (contours) of the
cluster
over-imposed onto the radio image
(grey scale). The X-ray brightness distribution shows structure,
with the inner region elongated in the NE-SW direction, and
the outer region elongated in a perpendicular direction (SE-NW)
with an extension to the West only seen in the X-ray emission.
The radio halo is located at the cluster center; its 
structure is elongated in the same direction as the central 
X-ray emission.

Proust \etal (2000) presented a study of the dynamics
of A520. The cluster seems to be undergoing 
strong dynamical evolution, since its cD galaxy 
it is not located at the center of the galaxy distribution.
Moreover they pointed out that this cluster may be
an example of a dynamically young system
where clumps of galaxies are still in phase of collapsing.

\subsection{A773}

In Fig. \ref{A773_a} we show
the radio images at 1.4 GHz of the cluster A773, with two
different angular resolutions.
The halo is quite regular, with a total  size of about 6$'$ (1.6
Mpc).
The brightness distribution of the extended emission is peaked in the center. 
However, in the VLA FIRST survey (Becker \etal 1995) no discrete source is
present in the central halo region at a level of 0.3 mJy/beam.

The higher resolution image  shows many discrete sources.
In particular it is evident that the elongated 
feature visible to the south in the low resolution map
is due to a blend of individual sources.

The  X-ray image (contours) presented in Fig. \ref{A773_b} 
is obtained with the ROSAT HRI,
by smoothing with a Gaussian filter of $\sigma=10''$.
The X-ray brightness distribution is very irregular
with a clear elongation, which is probably due
to the presence of two close peaks at the
distance of about 30$''$ (134 kpc). 
The radio halo is located at the cluster center,
with the brightest region coincident with the eastern X-ray peak.
The presence of X-ray substructures in A773 was also noted by
Rizza \etal (1998). 

This cluster has also been target for the study of the Sunyaev-Zeldovich
effect.
The temperature decrement detected toward A773 at 28.7 GHz is
$-575<\Delta T<-760 \mu$K (Carlstrom \etal 1996).

\subsection{A1664}

The 1.4 GHz radio image presented 
in Fig. \ref{A1664_a} (left panel) shows the diffuse source in detail.
The source is located at the projected distance of about
6$'$ (1.08 Mpc) from the cluster center, and is therefore a relic. 
From this image we derive a flux density of 50.2 mJy.
Due to the missing short spacings, some flux may be missed 
in our image.
This is confirmed in the NVSS image (see Giovannini \etal 1999
and Fig. \ref{A1664_a}, right panel)
where the structure appears much more extended, quite uniform
and the calculated flux density is 107 mJy.
No spectral index information is available  for this cluster, therefore
in the equipartition magnetic field strength and total power calculations 
we assumed a mean spectral index $\alpha=1.2$, as a typical value. 

\begin{figure}
\resizebox{9 cm}{!}{\includegraphics {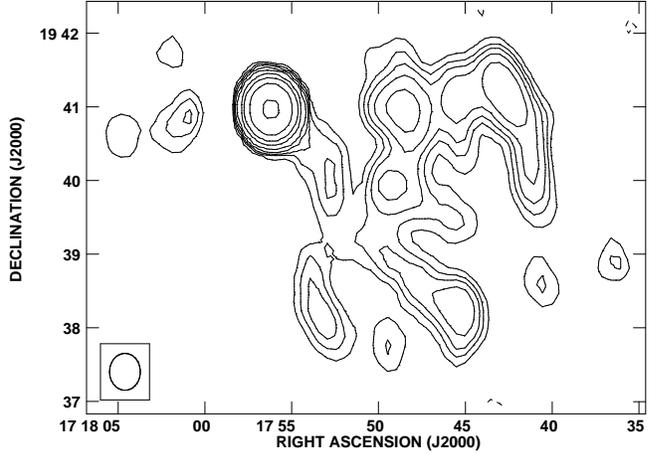}}
\hfill
\caption[]{Radio image at 1.4 GHz of the halo source in A2254.
 The FWHM is 30\arcsec$\times$25\arcsec;
the noise level is 0.1 mJy/beam. Contour levels are:
-0.3 0.3 0.5 0.7 1 1.5 3 5 10 30 mJy/beam.}
\label{A2254_a}
\end{figure}

\begin{figure}
\resizebox{9 cm}{!}{\includegraphics {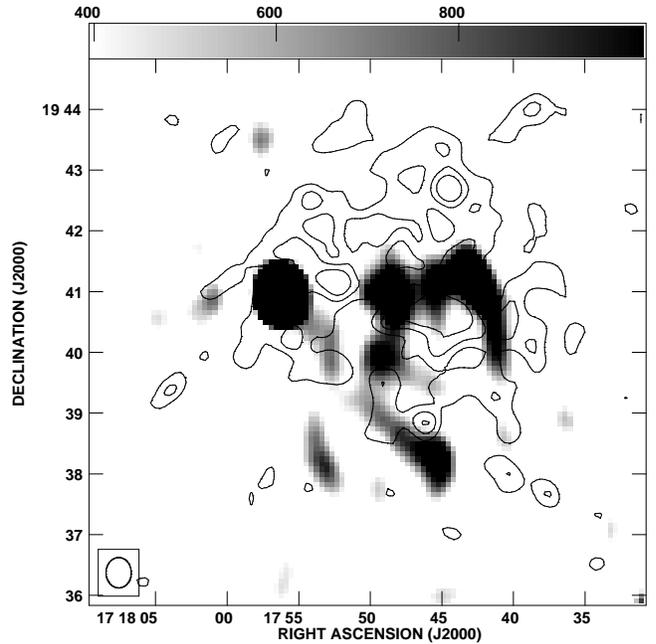}}
\caption[]{Isophotal contour plots
of the  X-ray (HRI) image taken from
the ROSAT archive overlapped to the grey-scale
image at 1.4 GHz 
(FWHM 30 \arcsec $\times$ 25\arcsec)
 of the cluster A2254.
The X-ray contours are: 1.1 1.3 1.5 2 2.5 
Counts/pixel (1pixel=5\arcsec$\times$5\arcsec).}
\label{A2254_bb}
\end{figure}

The ROSAT PSPC image (contours) for the cluster A1664, 
overlaid onto the radio image (grey-scale), is shown in Fig. \ref{A1664_b}.
The X-ray image was smoothed with a Gaussian of FWHM=30$''$.
The X-ray emission is asymmetric with an elongation toward
the south direction, where the relic is located.
Close to the X-ray emission center
there is a strong radio source well visible in the image.
This is identified with an optically line-luminous galaxy, 
studied by Allen \etal (1995).
Other radio galaxies are located inside and to the North of the relic.
This cluster shows a strong cooling flow, with 
a mass accretion rate of 260 M$_{\odot}$ year$^{-1}$ (Allen \etal 1995).

\begin{figure*}
\resizebox{18 cm}{!}{\includegraphics {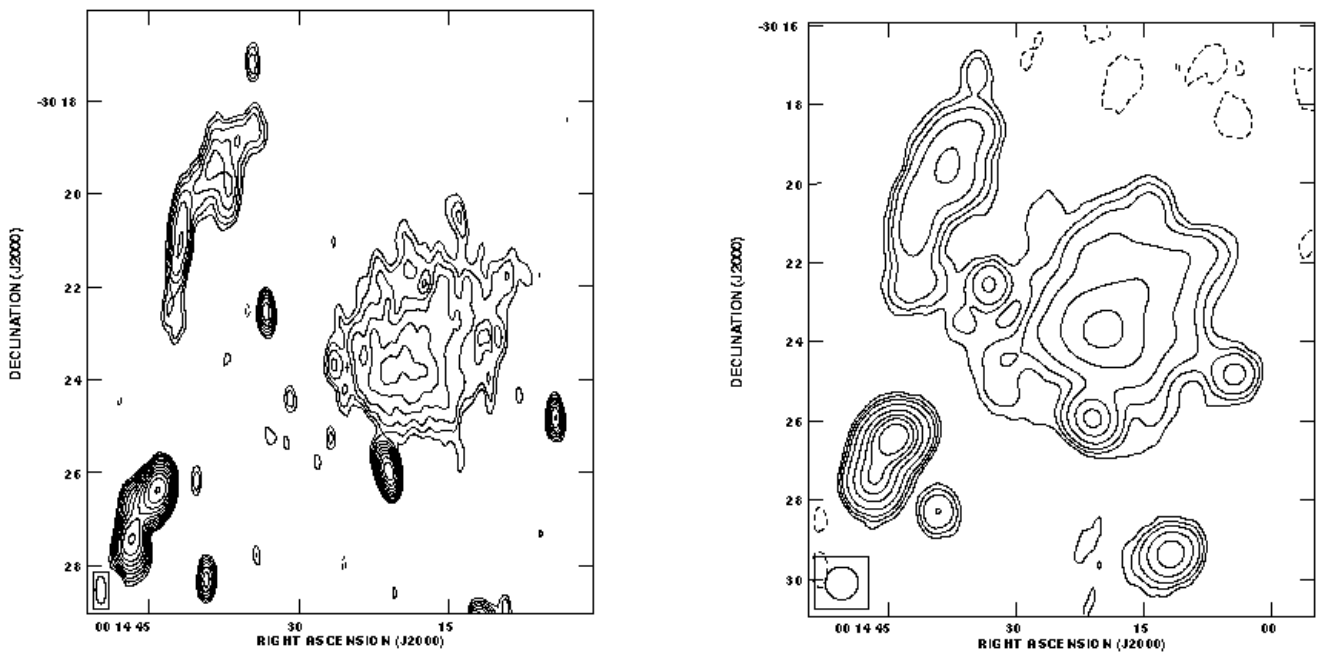}}
\caption[]{
Radio images at 1.4 GHz of the halo and the relic sources in A2744.
Left: the synthesized restoring beam is a Gaussian with FWHM of
36$''\times13''$.
The noise level is 0.04 mJy/beam. Contour levels are:
-0.20 0.20 0.28 0.40 0.57 0.80 1.31 1.60 2.26 3.20 4.52 6.40 9.05 12.80
18.10 25.60 36.20 mJy/beam.
Right: the synthesized restoring beam is a circular Gaussian with FWHM of
50$''$.
The noise level is 0.09 mJy/beam. Contour levels are:
-0.3 0.3 0.5 0.8 1.5 3 5 7 10 15 20 25 30 mJy/beam.
The elongated emission in the North-East peripheral region is a relic
source.
}
\label{A2744_a}
\end{figure*}

\subsection{A2254}

In Fig. \ref{A2254_a} we show
the radio image at 1.4 GHz of the radio halo 
in the cluster A2254.

The source located at the
position RA(2000)=17$^h$17$^m$56$^s$, 
DEC(2000)=19\degrees 40\arcmin 58\arcsec, with a flux of about 38 mJy
is an unrelated discrete source, while
the diffuse radio emission is not 
due to a blend of individual sources.
The radio structure of the halo appears quite irregular
with a projected size of about 5$'$ (1.16 Mpc).
This radio halo was also observed by Owen \etal (1999).

In Fig. \ref{A2254_bb} 
we show the overlay between the radio (grey scale) and 
the X-ray emission (contour).

Both the radio and the X-ray emission show a very clumpy and 
an irregular structure.
As expected from numerical 
simulations (e.g Schindler  \& M\"uller 1993), the 
irregular X-ray structure of A2254 indicates
a dynamically young cluster where different units are currently merging.
It is interesting that this cluster is the most irregular cluster 
in the set, not only in the X-rays but also in the radio morphology.
Most halos are found in very massive and hot clusters which are clusters
with a very deep potential. Even if the cluster has a
large surface brightness, the currently merging clumps are surely
characterized by smaller mass units and shallower potentials. It is
therefore remarkable and surprising that significant extended radio
emission is detected in this cluster.

\subsection{A2744}

In Fig. \ref{A2744_a} we present
the 1.4 GHz radio images of the cluster A2744, at different 
angular resolutions.
The diffuse radio emission consists of a radio halo at the cluster center,
and an elongated  emission in the North-East peripheral
region,  classified as a cluster relic.
The radio halo is rather regular in shape, 
with a size of about 7$'$ (2.34 Mpc),
but its brightness
distribution is asymmetric, being more extended toward NW.
The relic is at about 6$'$ from the cluster center and
has a projected size of about 6$'$ (2.0 Mpc).
In the image at lower resolution the relic seems to be connected
to the halo through a low surface brightness bridge.

\begin{figure}
\resizebox{8 cm}{!}{\includegraphics {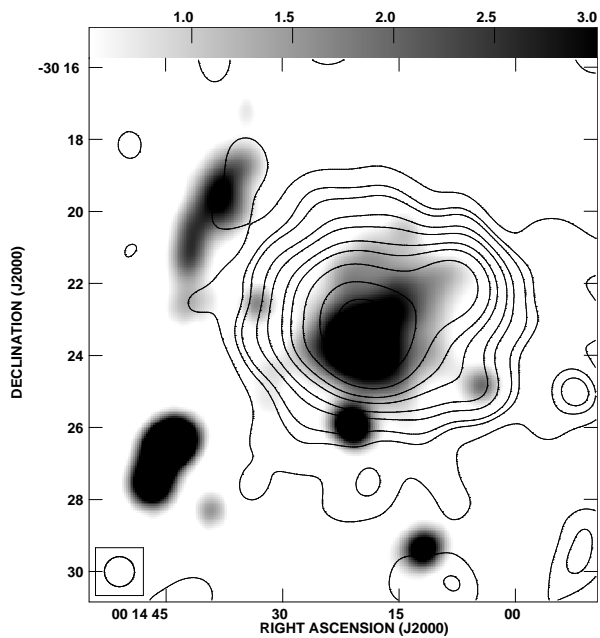}}
\hfill
\caption[]{Isophotal contour plots of the
 X-ray (PSPC) image taken from the ROSAT archive
overlapped to the grey-scale image at 1.4 GHz
(FWHM 50 \arcsec $\times$ 50 \arcsec) of the cluster A2744.
The X-ray contours are:
0.5 0.8 1 1.5 2 3 4 6 10 15 20 Counts/pixel (1pixel=15\arcsec$\times$15\arcsec).}
\label{A2744_b}
\end{figure}

X-ray data  for  A2744 are available from the ROSAT PSPC. 
In Fig. \ref{A2744_b} we present the X-ray image (contours)
obtained from the ROSAT archive by smoothing the
image with a Gaussian filter of $\sigma=30''$
over-imposed to the low resolution radio image (grey scale).
The X-ray emission shows substructure, which is a possible indication of a
recent merger.

The radio structure of the halo is extended in the same direction
as the X-ray emission.

Govoni \etal (2001) performed an analysis to quantify the similarity between the radio and the X-ray structure for A2744
and for three other clusters (Coma, A2255, A2319) containing
radio halos.
A linear relation between the radio and the X-ray brightness was 
found in A2744.

\section{X-Ray analysis}

A radial surface brightness profile was extracted from all the clusters.
For A520, A1664 and A2744 the radial profile was obtained 
by integrating the PSPC counts over concentric 
annuli of 15$''$, around the peak of the X-ray emission.
Similarly, the radial profile of the HRI X-ray emission 
in A115, A773, A2254 and A1664 was
obtained by integrating the photon counts over concentric
annuli of 5$''$.

\begin{figure*}
\vspace{19 cm}
\includegraphics{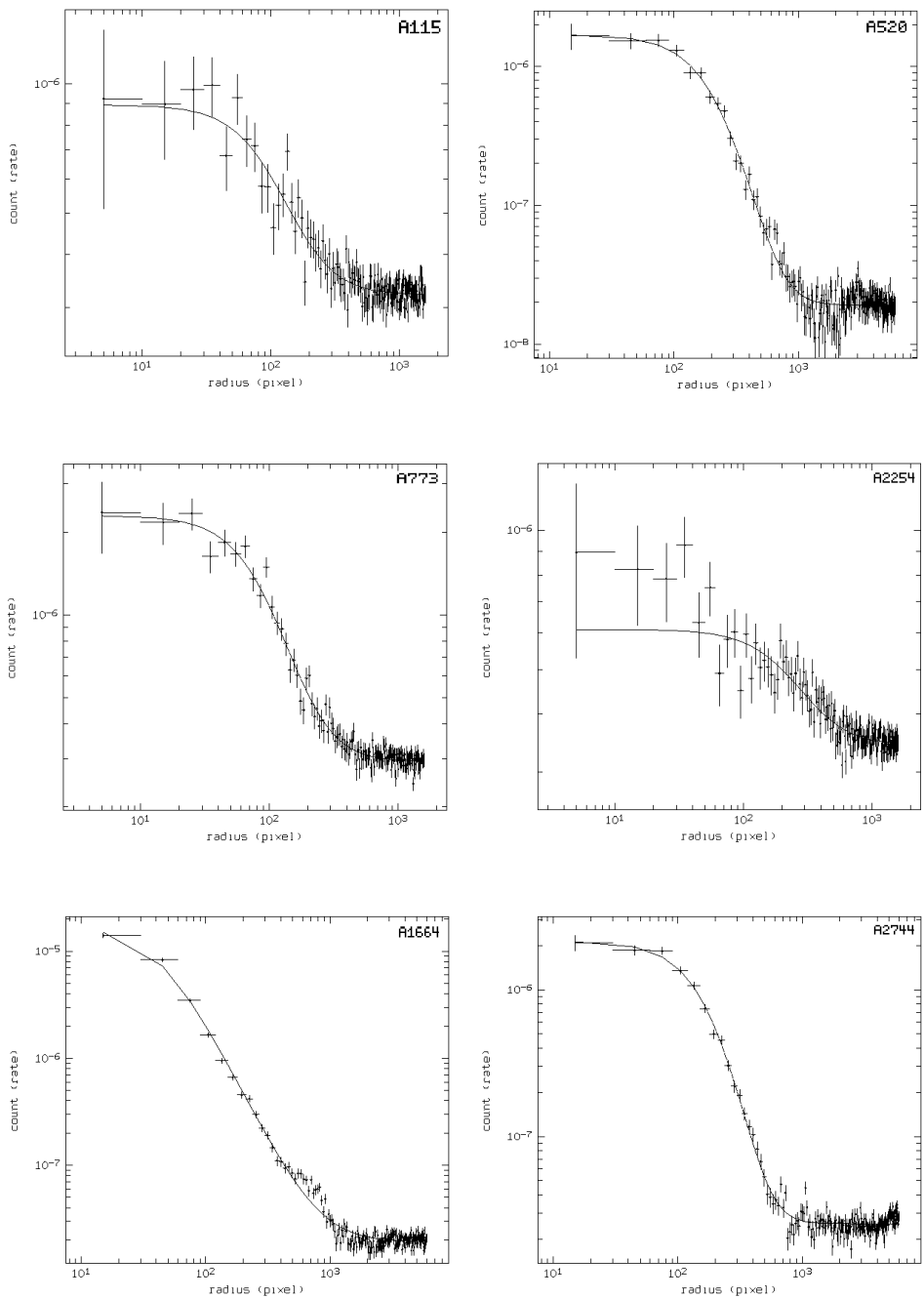}
\caption{
Surface brightness profile of the clusters. The full lines show the best fit
 of the $\beta$-model.}
\label{profili}
\end{figure*}

The profiles were fitted with a hydrostatic
isothermal $\beta$-model (Cavaliere \& Fusco-Femiano, 1981):
\begin{equation}
S(r)=S_0(1+r^2/r_{\rm c}^2) ^{-3\beta+0.5}+S_{\rm b}
\end{equation}
where S$_0$ is the central surface brightness, r$_{\rm c}$ is the core radius,
and $\beta$ is the ratio of the galaxy to the gas temperature.
The cluster background $S_{\rm b}$ was also fitted.
The azimuthally averaged surface brightness profiles are 
presented in Fig.\ref{profili}. The full line in each figure
represents the best fit of the hydrostatic isothermal model.

The parameters of the gas distribution obtained by fitting the $\beta$-model
to the surface brightness profile are reported in Table 6. 
For both the PSPC and HRI images we have
attempted to take substructure into account by masking out of the fit 
the substructure regions, the contaminating point sources
and characterizing the supposedly symmetric main body of the cluster.
Despite our caution in the analysis of the X-ray brightness
profiles, the problem of the presence of the clumpy structure is present
in most of the analyzed clusters and it is very severe
in the case of A2254. For these reasons the $\beta$-model parameters for the cluster A2254 are not included 
in Table 6.

In general the fit of the HRI surface brightness profile could be 
problematic since the X-ray cluster emission could extend over the 
whole field of view of the detector
and the profile could not reach the background.
This problem seems not to be present 
in the analyzed clusters, moreover, in A1664 where both the PSPC 
and HRI data was analyzed, the comparison of the two data set 
shows that the data 
are consistent.\\
Count rates, energy fluxes and X-ray luminosities 
are reported in Table 7.
For the calculations of the X-ray fluxes and luminosities the cluster temperatures have been taken from the literature (when available)
or from the $L_{\rm X} -T$ relation (Markevitch 1998). 
A quasi-bolometric luminosity, $L_{\rm X,bol}$, has been calculated in the
source rest frame energy range 0.01--40.0\,keV (for the relevant range of
cluster gas temperatures at least 99\,\% of the flux is contained in this
energy range).

\begin{figure*}
\vspace{19 cm}
\includegraphics{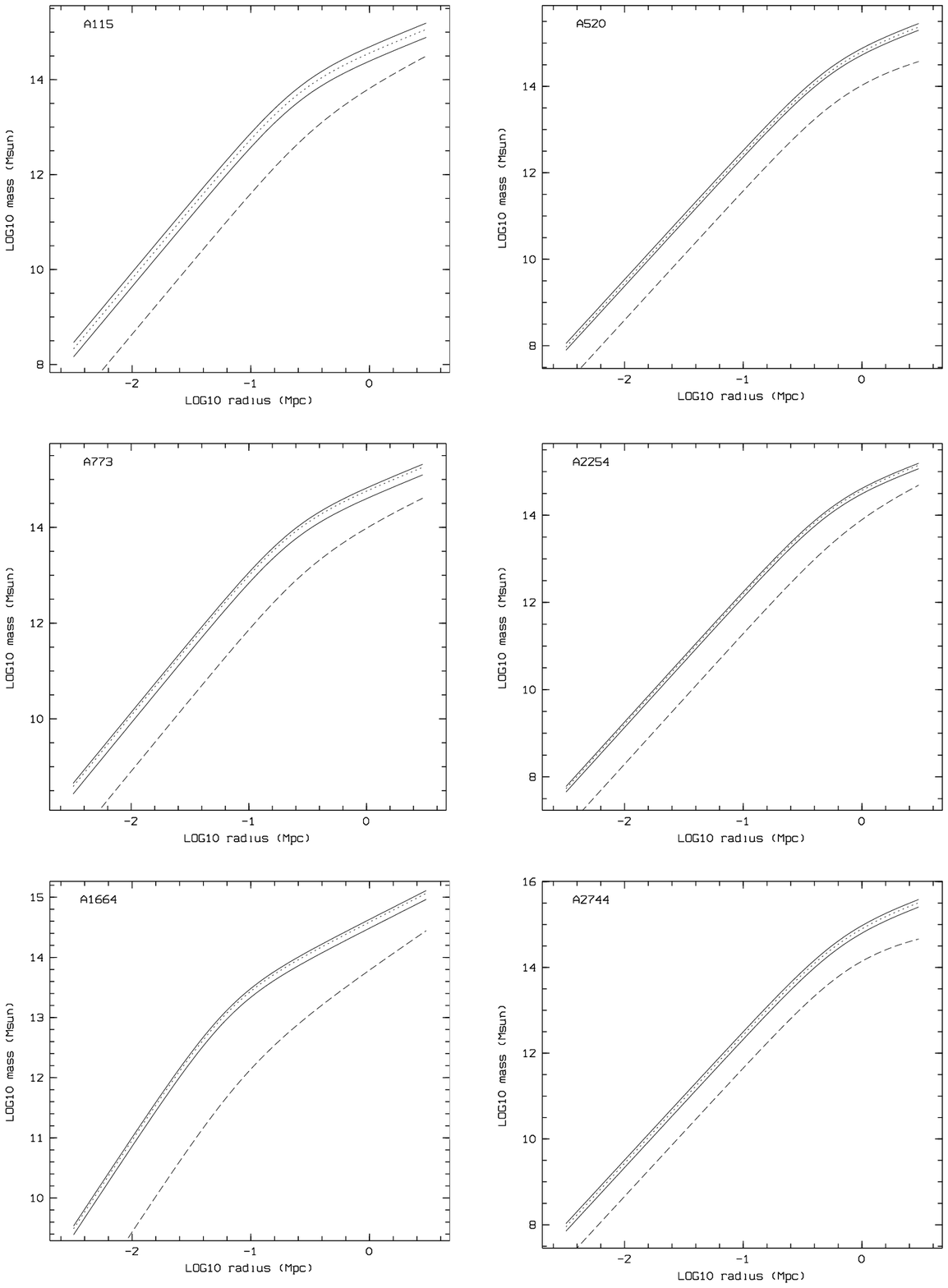}
\caption{
Mass profile of the clusters.
The dashed and dotted lines represent respectively the total gas mass
of the X-ray emitting gas, and the gravitating mass, obtained by
assuming hydrostatic equilibrium. Full lines around the dotted
line show the uncertainty
range for the gravitating mass due to the assumption of a temperature range. 
}
\label{mass}
\end{figure*}

Under the simple assumption of hydrostatic equilibrium in the ICM and
spherical symmetry of the cluster, it is possible to determine the
total cluster mass from the gas density and temperature distribution.
To determine the cluster mass, we use the gas density profile 
obtained from the $\beta$-model fit.
The intracluster gas has been assumed to be isothermal at the temperature
obtained from the literature.
A problem in the mass computation for clusters
possibly undergoing merger events could be
the assumption of isothermality and hydrostatic
equilibrium.
Simulations of Evrard et al. 1996 and Schindler 1996 show that
the mass determined in this way is in general reliable.
Observationally the study by Finoguenov et al. 2001 shows that the
assumption of isothermality in general may lead to an 
overestimate of the mass by about 20\% at the virial radius.

Since the luminous matter in clusters is dominated by 
the X-ray emitting gas, the fraction $M_{\rm GAS}/M_{\rm TOTAL}$ provides
a good measure of the amount of baryonic matter in the clusters. 
Gas fractions for nearby clusters range from about 
$\simeq$5 to 30 \% 
with a positively increasing 
radial gradient (B\"ohringer 1995, Evrard 1997). Fig. \ref{mass} shows 
the resulting mass profiles. The dashed and dotted lines represent 
the total gas mass of the X-ray emitting gas and the gravitating mass, 
respectively.
Full lines show the statistical range for the gravitating mass
calculated from the uncertainty in the temperature measurement.

When the inner regions of the clusters become very dense the 
cooling time of the gas may drop below the age of the 
system, and the cluster can form cooling flows.
We calculated the cooling time of each cluster, at the 
center, and established the existence of a cooling flow
when the cooling time was shorter than about $10^{10}$ years.

In the following the details of the X-ray analysis for each cluster are 
presented.

\begin{table*}
\caption{X-Ray fits}
\begin{flushleft}
\begin{tabular}{cccccc}
\hline
\noalign{\smallskip}
Name      &  S$_0$              & S$_{\rm b}$                 & $r_{\rm c}$      &$\beta$ & n$_0$   \\
          &Cts/s arcsec$^2$     & Cts/s arcsec$^2$      & arcsec &   & cm$^{-3}$        \\
\noalign{\smallskip}
\hline
\noalign{\smallskip}
 A115     &2.29$\times 10^{-06}$& 12.86$\times 10^{-07}$ & $58^{+37}_{-18}$ & 
$0.59^{+0.24}_{-0.11}$ & 3.6$\times 10^{-03}$ 
\\
 A520     &6.64$\times 10^{-06}$& 7.90$\times 10^{-08}$ & $127^{+17}_{-16}$ & 
$0.87^{+0.09}_{-0.07}$ & 3.2$\times 10^{-03}$
\\
 A773     &8.04$\times 10^{-06}$& 11.68$\times 10^{-07}$& $50^{+10}_{-7}$ & 
$0.63^{+0.07}_{-0.05}$
&6.7$\times 10^{-03}$ 
\\
 A1664    &6.89$\times 10^{-05}$& 7.65$\times 10^{-08}$ & $21^{+2}_{-2}$ & 
$0.55^{+0.01}_{-0.01}$ 
&22.8$\times10^{-03}$ \\
 
 A2254    &  -                  &         -              &  -  & -    
&           -        
\\ 
 A2744    &8.44$\times 10^{-06}$& 10.20$\times 10^{-08}$& $115^{+13}_{-10}$ 
& $1.00^{+0.09}_{-0.07}$
&3.8$\times 10^{-03}$

\\
\noalign{\smallskip}
\hline
\multicolumn{6}{l}{\scriptsize Col. 1: cluster name; Col. 2: central surface brightness; 
Col. 3: background surface brightness;}\\
\multicolumn{6}{l}{\scriptsize  Col. 4: core radius; Col. 5: $\beta$ parameter; Col. 6: central gas density.} 

\end{tabular}
\end{flushleft}
\end{table*}

\begin{table*}
\caption{X-Ray parameters}
\begin{flushleft}
\begin{tabular}{ccccc}
\hline
\noalign{\smallskip}
Name   &count rate   & flux(0.1-2.4 keV)      & L$_{\rm X}$ (0.1-2.4 keV) & L$_{\rm X,bol}$   \\
       & s$^{-1}$     & erg~s$^{-1}$~cm$^{-2}$ & erg~s$^{-1}$        & erg~s${^-1}$  \\
\noalign{\smallskip}
\hline
\noalign{\smallskip}
 A115     &$0.177\pm 0.008$& 9.57$\times 10^{-12}$ &  1.57$\times 10^{45}$  & 3.6$\times 10^{45}$
\\
 A520     &$0.309\pm 0.009$& 7.33$\times 10^{-12}$ &  1.22$\times 10^{45}$  & 3.4$\times 10^{45}$
\\
 A773     &$0.121\pm 0.004$ & 5.16$\times 10^{-12}$ &  1.01$\times 10^{45}$ & 2.9$\times 10^{45}$
\\
 A1664    &$0.429\pm 0.007$ & 1.05$\times 10^{-11}$ &  0.72$\times 10^{45}$ & 1.8$\times 10^{45}$           
\\ 
 A2254    &$0.105\pm 0.005$ &5.66$\times 10^{-12}$  &  0.76$\times 10^{45}$ & 1.9$\times 10^{45}$
\\ 
 A2744    &$0.265\pm 0.006$ & 5.38$\times 10^{-12}$ &  2.10$\times 10^{45}$ & 6.5$\times 10^{45}$
\\
\noalign{\smallskip}
\hline
\multicolumn{5}{l}{\scriptsize Col. 1: cluster name; Col. 2: X-ray count rate; Col. 3: X-ray flux, calculated in the 0.1-2.4 energy band;}\\ 
\multicolumn{5}{l}{\scriptsize Col. 4: X-ray luminosity, calculated in the 0.1-2.4 energy band; 
 Col. 5: X-ray bolometric luminosity.}\\ 

\end{tabular}
\end{flushleft}
\end{table*}

\subsection{A115}

The computation of the azimuthally averaged profile, was obtained
by masking the X-ray emission coming from the point source 3C28 
and by excluding the southern cluster sector corresponding 
to the substructure.
The centroid of the X-ray emission was taken approximately at the
position RA(2000) = 00$^h$ 55$^m$ 53.2$^s$, 
DEC(2000) =  +26\degrees 24\arcmin 59\arcsec. 
The parameters of 
the best fit $\beta$-model are given in Table 6.
In spite of  the exclusion of the contaminating regions,
the complex morphology of the cluster makes it difficult to fit 
the cluster with a $\beta$-model.
This distorted morphology and the different performed analysis 
could be the reason for the difference between our $\beta$-model parameters 
and the parameters calculated by Shibata \etal (1999).
They give the parameter values for the main peak of 
$\beta =1.05^{+0.08}_{-0.06}$ and r$_{\rm c}$=$1'.22_{-0.24} ^{+0.29}$.
Their ASCA observation shows a significant temperature
variation in this cluster, moreover a linking region 
between the main and the sub-cluster shows 
a high temperature compared with other regions, 
indicating that Abell 115 is a system in a stage of a merger event.
The ASCA spectrum of the main cluster is consistent with a cooling flow
model with a mass deposition rate of 89 M$_{\odot}$/yr with an upper limit of
419 M$_{\odot}$/yr. Our analysis leads to  a cooling time of the same 
order as the Hubble time. Therefore we conclude that a faint cooling 
flow with a low mass-rate seems to be present in the center of A115,
consistently with the ASCA results.
A cooling flow could be present in this cluster since the merger has 
not progressed enough to disturb the cluster center.

Ikebe \etal (2001 in prep.) 
give a temperature of 5.9 keV and 6.1 keV for the Northern and the Southern 
peak respectively.
These data are in agreement within the errors with the temperature data 
presented by Shibata \etal (1999).
For the estimate of the X-ray
luminosity, a temperature of 5.9 keV was used. 
The count rate reported in Table 7 
was calculated  excluding the source 3C28 but taking into account both
the substructures in the X-ray emission.

For the total cluster mass determination the gas temperature was 
taken in the range 4-8 keV.
The resultant uncertainty range for the gravitating mass profiles
is presented in Fig.~\ref{mass}, 
where the gas mass profile is also shown.
At a radius of 3 Mpc, the total mass of the cluster is in the range 
$0.77-1.5\times10^{15}$ M$_{\odot}$, 
and the gas mass fraction is respectively in the range 17-9\%
which is consistent with a typical gas mass fraction
in clusters.

\subsection{A520}

The X-ray emission in A520 has been traced out significantly 
to a maximum radius of 9$'$ (2.3 Mpc).

The radial profile was obtained using as a centroid the X-ray peak 
at position  RA(2000)=04$^h$54$^m$10.6$^s$, 
DEC(2000)=+02\degrees55\arcmin20\arcsec.
The parameters of the best fit of a hydrostatic isothermal model
are given in Table 6.

For the calculation of the X-ray
luminosity (see Table 7) a temperature of 8.3 keV was used 
(Ikebe et al. 2001, in prep.).

With the gas temperature in the range 7-10 keV,  
the total mass of the cluster, at a radius of 3 Mpc, is in the range 
$1.98-2.83\times10^{15}$  M$_{\odot}$, 
and the gas mass fraction is in the range 19-13\% (Fig. \ref{mass}).

Within a radius of 300 kpc the cooling time (0.34$\times$ 10$^{11}$ yr)
is found to be longer than the Hubble time, implying 
the absence of a cooling flow in this cluster, in agreement with
Allen \& Fabian (1998).
  
The absence of a cooling flow, the extended core radius and the 
high X-ray luminosity could indicate that A520 may be characterized
by a merger event.

\subsection{A773}

The X-ray emission of the cluster shows two peaks at a distance of
about 30$''$.

The surface brightness profile of the cluster was obtained without
 masking any region and by fixing the centroid 
of the X-ray emission between
the two peaks on the approximate symmetry center in the position 
RA(2000)=09$^h$17$^m$52.5$^s$, DEC(2000)=51\degrees43\arcmin41\arcsec.

The parameters of the $\beta$-model are given in Table 6.
These results are in agreement with the core radius and the $\beta$ 
obtained by Rizza \etal (1998).

To determine the total cluster mass the gas temperature 
was taken in the range 6-10 keV.
The total cluster mass profile and the gas mass profile 
of A773 are presented in Fig. \ref{mass}.
In the figure the dotted line represent the total cluster mass
considering the cluster in isothermal equilibrium 
with a temperature of 8.6 keV (Ikebe \etal 2001, in prep.).

\begin{figure}
\includegraphics[width=8cm, angle=-90]{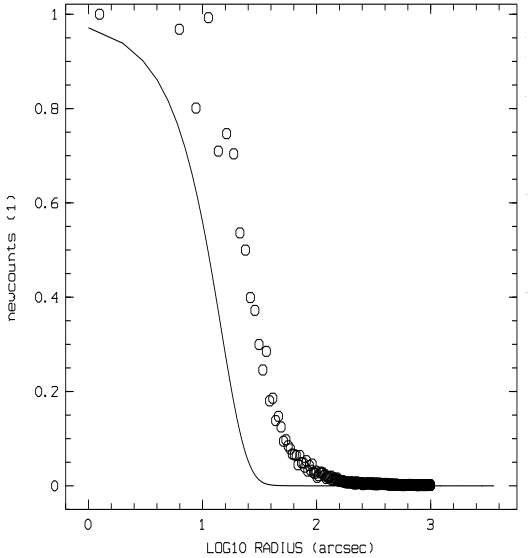}
\hfill
\caption[]{
Comparison of the normalized surface brightness profile (open dots)
of the cluster A1664 in the PSPC image with the point spread function (PSF)
of the PSPC detector (solid line).}
\label{A1664_psf}
\end{figure}
At a radius of 3 Mpc, the total mass of the cluster is in the range 
1.25 - 2.08 $\times 10^{15}$ M$_{\odot}$, 
and the gas mass fraction is in the range 32-19\%.  

Within a radius of 300 kpc the cooling time 
(0.17$\times$ 10$^{11}$ yr) is found to be only marginally larger
than the Hubble time, therefore we consider that this cluster 
has no significant cooling flow, 
in agreement with Allen \& Fabian (1998).

\subsection{A1664}

The X-ray emission of A1664 could be traced out significantly 
to a maximum radius of 11$'$ (2 Mpc).

The centroid of the radial profile was chosen on the maximum of the PSPC
image, i.e. at RA(2000)=13$^h$03$^m$43$^s$, 
DEC(2000)=$-$24\degrees14\arcmin45\arcsec. 
The fitting results are summarized in Table 6.

A1664 shows a very steep surface brightness profile
with a strong central peak.
The peaked central surface brightness could be due 
to a very strong cooling flow in the cluster center, or it
could be the result from a central point source.
To exclude the strong influence of a point source, Fig. \ref{A1664_psf} 
shows the comparison of the normalized surface brightness profile 
of the cluster in the PSPC image with the point spread function (PSF)
of the PSPC detector. From the figure it is clear that the central surface 
brightness peak is not caused by a point source.  

The total cluster mass was obtained taking the gas temperatures
in the range 5-8 keV.
The total cluster mass profile and the gas mass profile 
of A1664 are presented in Fig. \ref{mass}.
In the figure the dotted line represents the total cluster mass
considering the cluster in isothermal equilibrium 
with a temperature of 6.8 keV.
At a radius of 3 Mpc, the total mass of the cluster is in the range 
0.92-1.47 $\times 10^{15}$ M$_{\odot}$, 
and the gas mass fraction is in the range 30-19\%
which is consistent with a typical gas mass fraction
in cluster.  

Within a radius of 300 kpc the cooling time 
(0.44$\times$ 10$^{10}$ yr) was found to be smaller
than the Hubble time. This result and the presence of a strong central peak
in the surface brightness profile is consistent with the 
presence of a cooling flow in this cluster.

\begin{figure*}
\vspace{17 cm}
\includegraphics{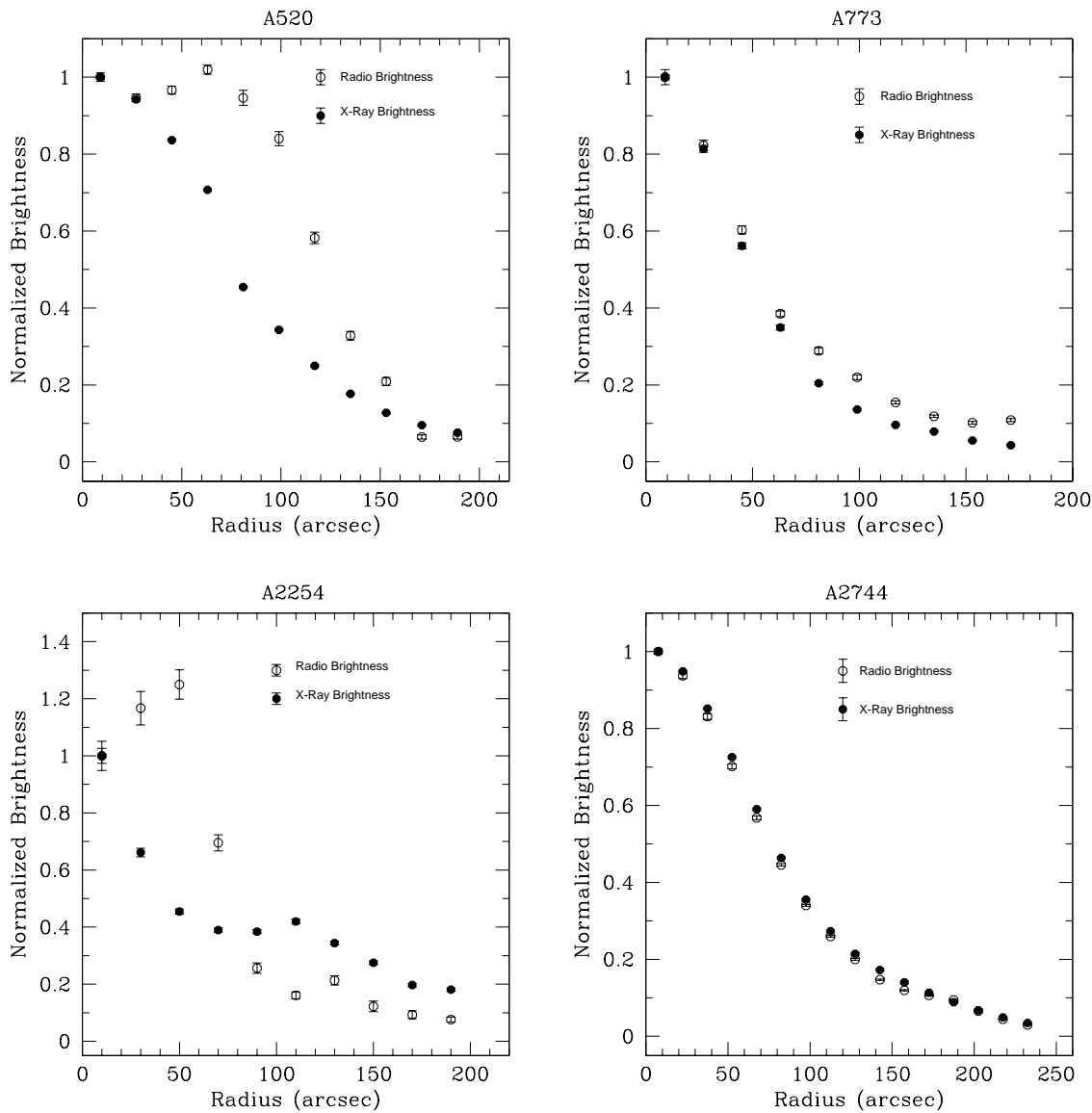}
\caption{ Normalized radio and X-ray profile of the four clusters 
containing a radio halo.}
\label{profiles}
\end{figure*}

\subsection {A2254}

The X-ray emission of A2254 is clumpy and irregular.
The surface brightness profile is therefore not well represented by a $\beta$-model.
 For this reason we do not give the best fit parameters.

The count rate reported in Table 7 was calculated excluding the X-ray point source visible on the East side of the X-ray image.

The total cluster mass was estimated with temperature in the range 6-8 keV,
and rough $\beta$ parameters.
These values should be taken cautiously, since the simple $\beta$-model appears
 not adequate to describe the observed X-ray brightness profile for this cluster. 
At a radius of 3 Mpc, the total mass of the cluster is in the range 
1.8-2.4 $\times 10^{15}$ M$_{\odot}$, 
and the gas mass fraction is in the range 21-16\% (see Fig. \ref{mass}) .  
 
The cooling time is found to be longer than the Hubble time, implying 
the absence of a cooling flow in this cluster.

\subsection {A2744}

The X-ray emission of A2744 has been traced out significantly 
to a maximum radius of 11$'$ (3.7 Mpc).
The X-ray peak is at RA(2000) = 
00$^h$ 14$^m$ 18.7$^s$, DEC(2000) = $-$30\degrees 23\arcmin 16\arcsec. 
The X-ray emission is characterized by the presence of substructure. 
In particular a sub-clump is located at a distance of 
about 150$''$ from the cluster center on the North-West side.

The derived azimuthally averaged surface brightness was obtained
by excluding the substructure on the north-west side.
The centroid of the X-ray emission was taken in coincidence
with the X-ray peak.
The surface brightness profile results quite regular
and well described by a $\beta$-model.
The parameters obtained from the fit are presented in Table 6.

Fig. \ref{mass} shows the gas mass profile and the total mass
profile calculated assuming an isothermal temperature
profile in the temperature range 8-12 keV. The dotted line represents 
 the total cluster mass calculated considering a temperature of 
 10.1 keV (Ikebe \etal 2001, in prep.).  
At a radius of 3 Mpc, the total mass of the cluster is in the range 
2.55-3.82 $\times 10^{15}$ M$_{\odot}$, 
and the gas mass fraction is in the range 18-12\%
which is consistent with a typical gas mass fraction
in cluster.  
 
In agreement with Allen \& Fabian (1998), 
the cooling time (0.31$\times$ 10$^{11}$ yr) of A2744
is found to be longer than the Hubble time, implying 
the absence of a cooling flow in this cluster.

Due to the absence of cooling flow and the presence of 
substructures A2744 could be a cluster with an ongoing merger event.

\section{Discussion and conclusions}

We confirm the presence of diffuse radio emissions 
in the six clusters analyzed here.
In A520, A773, A2254 and A2744 radio halos are detected at the cluster center. 
In A115, A1664 and A2744 a peripheral source is present. 
We note that both a halo and a relic are present in A2744,
 like in other cases known in literature 
(e.g. Coma, A2255, A2256).

The halos in A520, A773 and A2744,
appear extended and quite regular.
On the contrary, the halo in A2254 is very clumpy
and irregular. In this cluster, also the X-ray emission is quite irregular.
   
The presence of a diffuse radio emission in these clusters is a 
direct evidence for the presence of a magnetic field associated
with the intergalactic medium of the cluster.
Under equipartition conditions, 
we estimate magnetic fields of the order of 1 $\mu$G, permeating a large
region of the clusters, up to $\sim$ 1 Mpc radius.

From our data we can obtain some indications for the presence
of a merger process and cooling flow in the clusters. 
Clumpy and unrelaxed X-ray structures can be considered
as indicators of cluster mergers.
In particular, the clusters A2744 and A773 could be
in the process of an ongoing merger while
A520 may be at a later merger state.
Finally, the clumpy
structure in A2254 could indicate a dynamically young cluster
where the merging involves several different small units.
Therefore, we confirm that the formation of the halos 
sources could be connected to the presence
of a cluster merger process, although the stage of the merger process
can be very different. 
We note that the clusters A2744, A520 A773 and A2254
are all non-cooling flow clusters. 
This is expected, since simulations (McGlynn \& Fabian 1984, 
Roettiger et al. 1996, G\'omez et al. 2001) suggest
that mergers can disrupt cooling flows.
 
While until now there is no evidence of clusters with both
radio halo and massive cooling flow, 
clusters with a cooling 
flow can host a relic source as A85 or A133.
The cluster A1664 is another example.
The distorted X-ray image of A1664 could indicate 
the presence of a recent merger, probably between
the main body of A1664 and a smaller unit.
This  produced an asymmetric
outgoing shock wave and the disturbance did not reach the cluster
center with enough effect to disrupt the cooling flow. 
This is in agreement with the current interpretation of 
a relic source as the product of an outgoing merger shock wave 
(En{\ss}lin \etal 1998, Roettiger \etal 1999).

Also in the cluster A115 we have evidence of a merger in a early stage 
and consistent with the presence of a faint cooling flow. In this cluster
as discussed before a relic source have been detected.

From previous studies,
some clusters were found to show 
a close morphological similarity 
between the radio and the X-ray emission (e.g. Coma, A2255, A2319).

In Fig. \ref{profiles} we compare the shape of the normalized
radio and X-ray brightness profile of the four clusters
containing a radio halo. To obtain these profiles, the mean 
and the mean error were estimated within concentric rings centered
on the X-ray symmetry center of the clusters.
Point sources were excluded from the analysis.
As already noted in Govoni \etal (2001), in A2744 
we found a perfect similarity of the two profiles.

In A773 the two profiles are quite similar, while in A2254 and in A520
the radio and X-ray profiles show a different shape.
In A2254, the different profiles reflect the fact that this cluster 
is very clumpy in both the radio and X-ray emission.
In A520, the halo is elongated, whereas the X-ray emission is much more
extended and symmetric.
Since the A520 radio profile
appears flat, we cannot exclude that this source 
is not a real halo but a source at the cluster periphery
and seen in projection onto the cluster center. This possibility
should be checked in future also with polarization studies.

\begin{figure}
\includegraphics[width=8cm]{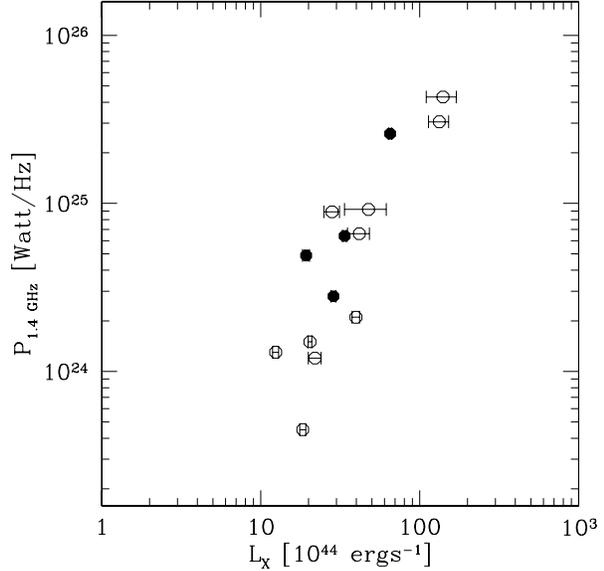}
\hfill
\caption[]{
Relation between the halo radio power at 1.4~GHz and cluster X-ray bolometric luminosity. 
 Filled and open dots represent the clusters studied in this work and the clusters
 taken from the literature, respectively.
}
\label{xlumin}
\end{figure}

\begin{figure}
\includegraphics[width=8cm]{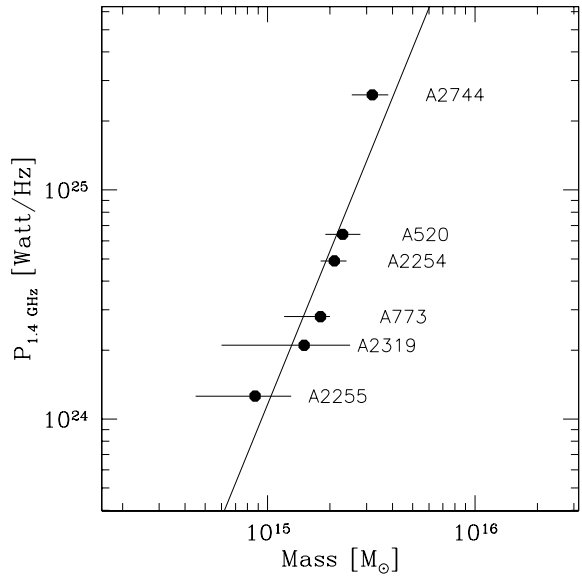}
\hfill
\caption[]{
Relation between the halo radio power at 1.4~GHz and the gravitational mass 
for the clusters studied in this work, A2319 (Feretti \etal 1997a) and A2255 
(Feretti \etal 1997b).
}
\label{massrel}
\end{figure}

The radio properties of halos are linked to the properties
of the X-ray emitting gas through a correlation between the halo
monochromatic radio power and the X-ray luminosity (Liang et al. 2000,
Feretti 2001). 
In Fig \ref{xlumin} we present the relation,
with the addition of the new data obtained in this work.

We checked if the correlation is real or induced by distance effects. 
Since it is still present plotting the radio flux versus the X-ray flux
and considering that the slope of the correlation
is steep, we conclude that it is real.

The clusters hosting radio halos analyzed in this work are all
X-ray luminous and consequently they have a high temperature and
large mass. 

Since the cluster temperature (e.g., Markevitch et al. 1998)
 and the cluster mass (e.g., Reiprich \& B\"ohringer 1999) 
are related to the X-ray luminosity, a relation is expected 
also between the
monochromatic radio power and the cluster temperature (Liang et al. 2000)
and between the monochromatic radio power and the cluster mass.   

In all the analyzed clusters, mass values larger 
than 1.1$\times$$10^{15}$ M$_{\odot}$ are found for the total 
gravitating mass within 3 Mpc.
In particular, the two clusters A115 and A1664 hosting relic
sources have a total mass of about 1.1$\times$$10^{15}$ and 
1.2$\times$$10^{15}$ M$_{\odot}$, respectively,
while the mass of the clusters with radio halos range between 1.8$-$3.2
$\times$$10^{15}$ M$_{\odot}$. The most massive cluster is A2744 
which hosts both a relic and a halo source.

In Fig. \ref{massrel} we plot the radio power calculated at 1.4 GHz versus 
the gravitating mass (within 3 Mpc). 
There is a clear trend of increasing radio power for increasing mass.
The relation is well fitted with a 
power law $P_{\rm 1.4GHz}\propto{\rm Mass}^{2.2}$.

All the points in the fit are
from this work except for A2255 and A2319
whose values are taken from previous works (Feretti et al 1997a; Feretti
et al. 1997b) where the mass determination was performed in the same 
way as in this work.

As discussed above this relation reflects the already known 
relation between the radio power
and X-ray luminosity, but the trend shows a small scatter.
If confirmed, this trend could indicate that the real connection between 
the radio and the X-ray properties of the gas is primarily
related to the cluster mass rather than to the cluster temperature.

The cluster mass is a logical candidate for a fundamental
parameter since the energy available to accelerate
relativistic particles in a merger scales as $\simeq M^{2}$
as discussed by Buote (2001).
The existence of halos in massive clusters is also
consistent with their serendipitous detection
in clusters observed during the attempts to detect
Sunyaev-Zeldovich effect in massive high redshift
clusters (Liang et al. 2000, Feretti et al. 2000). 

We wish to point out that not all X-ray luminous clusters 
show a radio halo. Indeed, diffuse halos are present
in about 30-35\% of clusters with L$_{\rm X}>10^{45}$ erg s$^{-1}$
(e.g. Feretti et al. 2000).  
Therefore, the large mass (high X-ray luminosity) of a cluster
is not the only condition that ensures the existence of diffuse radio
emission. Also the presence of a recent merger event in the
cluster is needed for the formation and maintenance
of a radio halo, as stressed by Feretti (2001) and Buote (2001).
These results are in line with the recent model proposed
by Brunetti et al. (2001) where the importance
of the cluster dynamical history and of a recent merger event is
pointed out.  

\section*{Acknowledgments}

F.G. acknowledges the MPE of Garching for hospitality and partial 
financial support. 
We thank the anonymous referee for helpful comments.
NRAO is a facility of the National Science Foundation, 
operated under cooperative
agreement by Associated Universities, Inc. 
This work was partly supported by the Italian Ministry for University
and Research (MURST) and by the Italian Space Agency (ASI).
This research has 
made use of the
NASA/IPAC Extragalactic Data Base (NED) which is operated by the JPL, 
California Institute of Technology, under contract with the National 
Aeronautics and Space Administration.

\end{document}